# NbN superconducting nanowire single photon detector with efficiency over 90% at 1550 nm wavelength operational at compact cryocooler temperature

W. J. Zhang, L. X. You*, H. Li, J. Huang, C. L. Lv, L. Zhang, X. Y. Liu, J. J. Wu, Z. Wang, and X. M. Xie

State Key Lab of Functional Materials for Informatics,

Shanghai Institute of Microsystem and Information Technology (SIMIT), Chinese Academy of Sciences (CAS), Shanghai, 200050, P. R. China

CAS Center for Excellence in Superconducting Electronics (CENSE)

865 Changning Rd., Shanghai, 200050, P. R. China.

*lxyou@mail.sim.ac.cn

## Abstract

The rapid development of superconducting nanowire single-photon detectors (SNSPDs) over the past decade has led to numerous advances in quantum information technology. The record for the best system detection efficiency (SDE) at an incident photon wavelength of 1550 nm is 93%. This performance was attained from an SNSPD made of amorphous WSi; such SNSPDs are usually operated at sub-Kelvin temperatures. In this study, we fabricated an SNSPD using polycrystalline NbN. Its SDE is 90.2% at 2.1 K for incident photons with a 1550-nm wavelength, and this temperature is accessible with a compact cryocooler. The SDE saturated at 92.1% when the temperature was lowered to 1.8 K. We expect the results lighten the practical and high performance SNSPD to quantum information and other high-end applications.

## Main text

A single-photon detector with high detection efficiency is the key enabling technology for quantum information and various applications, including the test of loophole-free Bell inequality violation[1], quantum teleportation[2], measurement-device-independent quantum key distribution[3], and linear optical quantum computation[4]. Superconducting single-photon detectors outperform their semiconducting counterparts in terms of not only detection efficiency but also dark count rate, timing jitter, and counting rate[5]. In the case of the telecommunication wavelength (1550 nm), the highest system detection efficiency (SDE) greater than 90% has been reported for two types of detectors. One is a transition edge sensor (TES) made of tungsten (W), with an SDE of 95%[6]; the other is a superconducting nanowire single-photon detector (SNSPD) made of amorphous WSi, with an SDE of 93%[7]. However, because of the low superconducting transition temperature of W and WSi, the requirement of sub-Kelvin cryogenics represents a burden for practical applications. Many studies focused on SNSPDs fabricated using different materials and aiming to obtain a high SDE at higher operating temperatures have been reported[8-11]; however, none of these attempts has been successful. Regarding another important parameter, timing jitter, a WSi SNSPD and a W TES have values of approximately 150 ps and 50–100 ns, respectively, which limits their use in applications that require lower timing jitter.

The SDE of SNSPD ($\eta_{SDE}$) can be expressed as the product of three contributions[7]: $\eta_{SDE} = \eta_{oc} \times \eta_{abs} \times \eta_{int}$, where $\eta_{oc}$ is the optical coupling efficiency between the incident photons and the active area, $\eta_{abs}$ is the absorption efficiency of the nanowires, and $\eta_{int}$ is the intrinsic detection efficiency (IDE), which describes the pulse generation probability of the nanowire when a photon is absorbed. To achieve a high SDE, all three parameters must be simultaneously maximized. The $\eta_{oc}$ is mainly governed by the size of the active area of the SNSPD when the optical alignment method is fixed. The factors that influence $\eta_{abs}$ include the geometric and optical parameters of the nanowire (e.g., thickness, width, filling ratio, and refractive index) and the optical absorption enhancement structures (i.e., the cavity structures). The $\eta_{int}$ is mainly determined by the superconductivity of the nanowire; however, the factors are complicated[12-14]. On the one hand, geometric parameters such as the width and thickness of the nanowire play important roles. On the other hand, the operating parameters (e.g., bias current and operating temperature) can control the $\eta_{int}$ directly. In addition, good uniformity of the nanowire with respect to thickness variation and physical constrictions handicaps $\eta_{int}$ because it suppresses the maximum bias current. The aforementioned analysis indicates that $\eta_{abs}$ and $\eta_{int}$ are not independent; they are both related to the thickness and width of the nanowire. The correlation between $\eta_{abs}$ and $\eta_{int}$ makes the improvement of SDE complicated. Indeed, a WSi SNSPD with a high SDE is relatively simple because $\eta_{int}$ can easily reach 100% with a wide range of nanowire thicknesses and widths because of the lower pair-breaking energy of WSi. By contrast, in the case of an NbN SNSPD, the pair-breaking energy of NbN is nearly double that of WSi. As a result, the room for tuning $\eta_{abs}$ and $\eta_{int}$ is highly limited, which makes it challenging to achieve a high SDE.

NbN SNSPDs are a promising choice for practical applications because of their high critical temperature, which introduces the possibility of using a compact cryocooler for operation. Unfortunately, the highest SDE at 1550 nm wavelength for an NbN SNSPD is less than 80%[15-17]. Suspicion exists about the suitability of NbN as a material for fabricating an SNSPD with an SDE greater than 90%.

In this manuscript, we eliminate the aforementioned suspicion by demonstrating an NbN SNSPD with an SDE greater than 90% at 1550 nm. We introduced a distributed Bragg reflector (DBR) mirror onto a Si substrate; the measured reflectance of the mirror was greater than 99.9%, and the calculated absorption of the nanowire was greater than 99%. Excellent front-side single-mode lens fiber coupling was ensured by a large sensitive area (greater than ⌀15 μm). By tuning the film thickness, we maximized the $\eta_{abs}$ and $\eta_{int}$ simultaneously. As measured using a dilution refrigerator, the saturated SDE reached 92.1% when the temperature was less than 1.8 K. At 2.1 K, the measured SDE was 90.2% at a dark count rate (DCR) of 10 Hz; we subsequently verified this SDE by cooling the SNSPD using a compacted Gifford–McMahon (G–M) cryocooler. In addition, because of the high switching current of the NbN SNSPD, a relatively small timing jitter of 79 ps was achieved.

**Results**

**Device design and structure.** The design of the SNSPDs that operate at incident photon wavelength of 1550 nm was based on the principle of integrating a DBR mirror into the device to enhance the absorption of the nanowires. As shown in Fig. 1(a), the DBR comprised fifteen periodic $SiO_2/Ta_2O_5$ bilayers. The periodic bilayers were alternately deposited onto a 400-μm-thick Si wafer using ion-beam sputtering. Their thicknesses were fabricated to be one-fourth of the 1550-nm incident photon wavelength, i.e., 265 nm for the $SiO_2$ layer ($n_{SiO2}$ = 1.46) and 180 nm for the $Ta_2O_5$ layer ($n_{Ta2O5}$ = 2.15). The root-mean-square (RMS) roughness of both the DBR substrate and the deposited NbN thin film on the DBR was less than 0.3 nm (Supplementary Fig. 1). The measured reflectivity of the DBR substrate was greater than 99.9% at 1550 nm (Supplementary Fig. 2).

The nanowire linewidth was empirically designed to be 75 nm to enhance its $\eta_{int}$. Before fabricating the devices, we systematically conducted numerical simulations in which parameters such as the filling factors, thickness, and configuration of the cavity were varied while the wavelength remained fixed (Supplementary Fig. 3). The results of the simulations indicated that a maximum absorption efficiency greater 99% can be realized with an NbN thickness of approximately 7 nm and a filling factor of approximately 0.5 when the NbN film is integrated into a half-cavity (DBR/NbN/Air).

To experimentally validate the simulation results, we prepared devices with various filling factors (0.58–0.33) and different film thicknesses (6.5–8.0 nm) covering two active areas (⌀15 and ⌀18 μm). Scanning electron microscopy (SEM) images of the nanowire with a nominal 75-nm width and 140-nm pitch are shown in Fig. 1(b); these images demonstrate good controllability of the fabrication process. Figure 1(c) shows a transmission electron microscopy (TEM) image of the DBR mirror. The measured thicknesses of the deposited bilayers were observed to vary by ±4 nm from the designed values. Figure 1(d) shows a high-magnification TEM image of a cross-section of a nominal 7-nm-thick, 75-nm-wide nanowire, which has a slightly trapezoidal shape because of the side etching. The TEM image indicates an estimated thickness of approximately 8 nm, which includes an oxidized layer (1- to 2-nm thick) on the top[18].

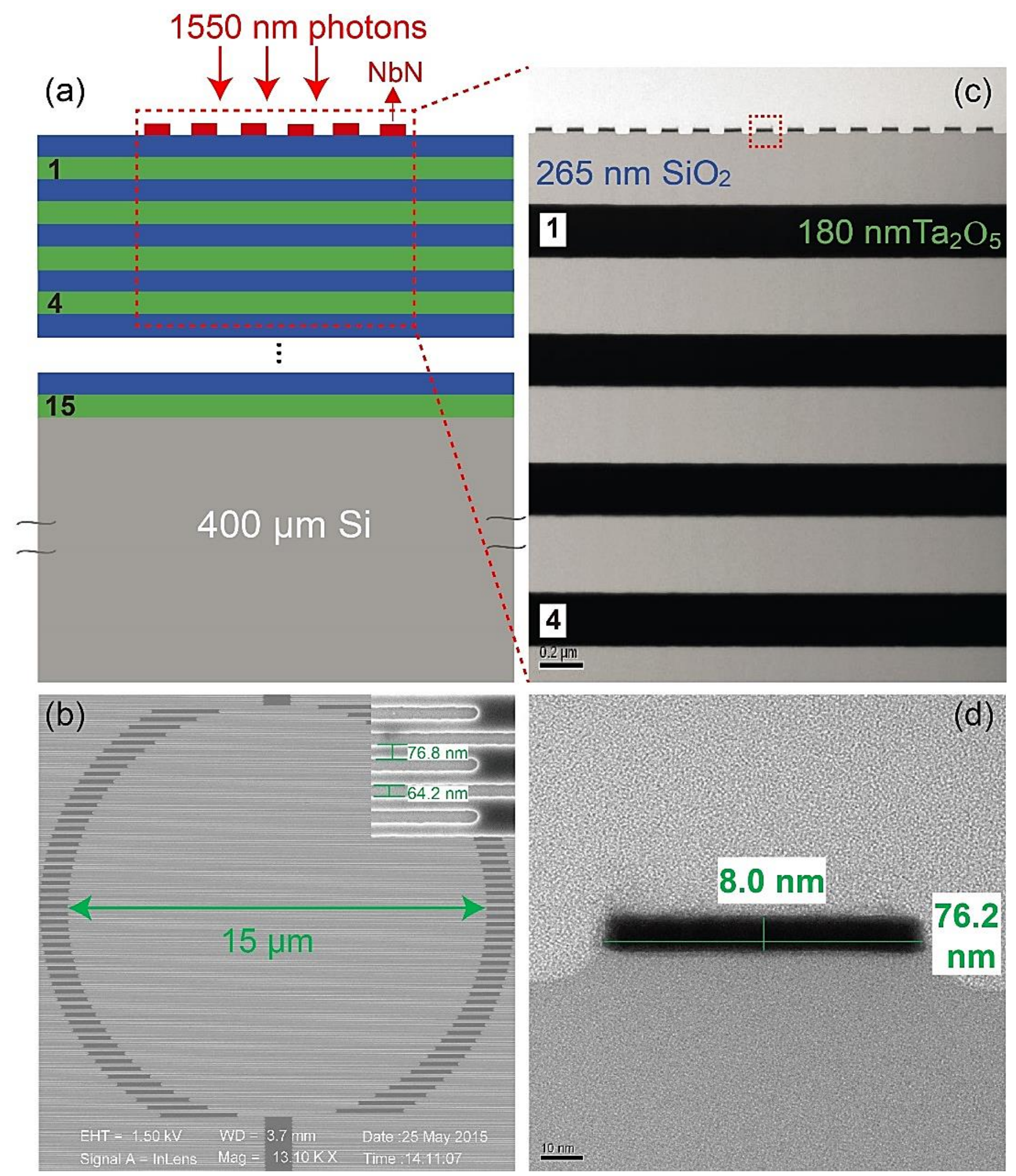

Fig.1. (Color online) (a) Schematic of a 7-nm-thick NbN SNSPD fabricated on a DBR/Si substrate. The DBR substrate comprised fifteen periodic $Ta_2O_5/SiO_2$ bilayers deposited onto a Si wafer. (b) SEM image of the deposited nanowires. The diameter of the active area was 15 μm. The inset shows a high-magnification image of the nanowires, which exhibit a nominal 77-nm width and 64-nm spacing. (c) TEM image of a cross section of nanowires on the DBR mirror. The thicknesses of the $SiO_2$ and $Ta_2O_5$ layer were approximately 265 and 180 nm, respectively. (d) High-magnification TEM image of the cross section of a single nanowire, whose measured thickness and width were approximately 8.0 and 76 nm, respectively. Because an existing oxidization layer with a thickness of approximately 1–2 nm was present on the top, the effective NbN thickness of the nanowire was estimated to be 7 nm.

**Device Characterization.** We characterized different SNSPDs with four film thicknesses (6.5, 7.0, 7.5, and 8.0 nm) fabricated in the same process run (Supplementary Table 1). The corresponding $T_c$s were 7.8–8.6 K and increased monotonically with increasing film thickness (Supplementary Fig. 4). We screened the detectors by selecting those with the highest switching current for the same geometrical parameters. The selected devices were then optically aligned to lens-single-mode fibers and mounted in a cryogen-free dilution refrigerator (CDR) with a wide temperature operating range and capable of reaching 16 mK.

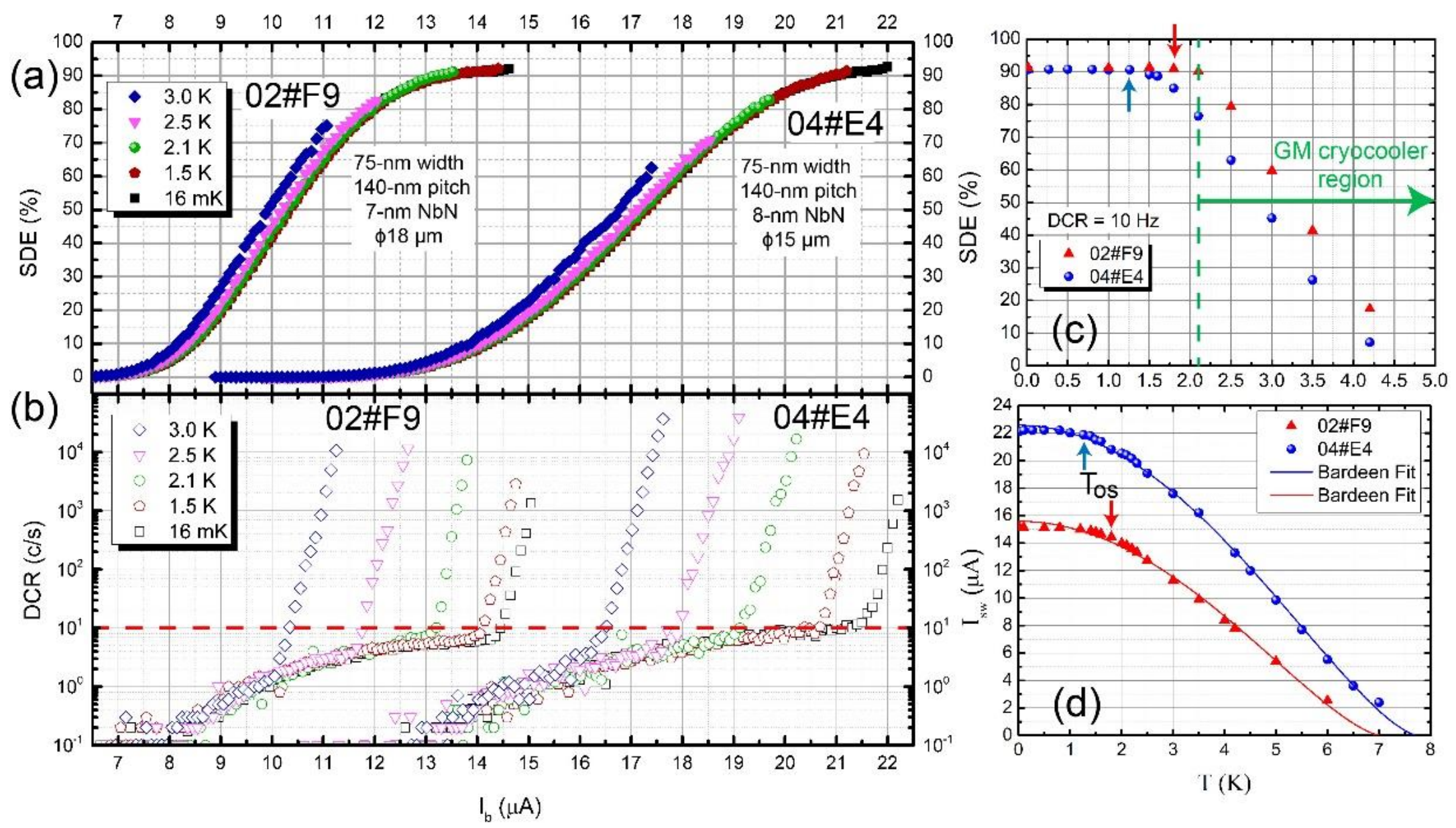

Fig. 2. (Color online) (a) and (b) SDEs and DCRs as functions of the bias current $I_b$ for the 7- and 8-nm-thick, 75-nm-wide, 140-nm-pitch SNSPDs, as measured from 3 K to 16 mK. The filled and open symbols represent the SDE and the DCR, respectively. As the temperature was decreased, a clear saturation plateau emerged. Thinner NbN films resulted in wider saturation plateaus. The maximal SDEs of devices 02#F9 and 04#E4 were 92.1% and 91.7%, respectively, when the devices were operated at a DCR of 10 Hz. (c) Temperature dependence of SDEs for devices 02#F9 and 04#E4 when the DCR was set to 10 Hz. For 02#F9, the SDE of 90.2% was obtained at 2.1 K (indicated by the green-dashed line), which is also the base temperature for a G–M cryocooler. The arrows indicate the onset temperature ($T_{os}$), where the SDE nearly saturated with decreasing temperature. (d) The switching current $I_{sw}$ as a function of temperature for devices 02#F9 and 04#E4.

Figures 2(a) and 2(b) shows the SDEs for an incident photon wavelength of 1550 nm and the DCRs as functions of the bias current $I_b$ for two devices, 02#F9 (7-nm thick) and 04#E4 (8-nm thick), at various temperatures. For clarity, five curves at five temperatures (3.0, 2.5, 2.1, 1.5 K, and 16 mK) are displayed. In the DCR measurement, the optical-fiber port at room temperature was shielded to eliminate any stray light from the environment. As the operating temperature was decreased, the SDE increased and a saturation plateau appeared. When the temperature was less than 3 K, all the SDE($I_b$) curves nearly overlapped at the same bias current. At 1.5 K and lower, we observed the highest SDE value of 92.1% for device 02#F9 at a DCR of 10 Hz and saturation plateau indicating that the $\eta_{int}$ reached approximately 100%. When the temperature was increased to 2.1 K (the lowest temperature that can be achieved with a commercial compact G–M cryocooler), an SDE of 90.2% was obtained with a DCR of 10 Hz. The 8-nm-thick NbN SNSPD (04#E4) exhibited a larger switching current ($I_{sw}$) of 22.5 μA compared to device 02#F9, whose $I_{sw}$ was 15.2 μA; the highest SDE of this device (91.7%) was measured at a DCR of 10 Hz and at 16 mK. However, the SDE decreased to 76.5% at 2.1 K because of the weak saturation of SDE. To further verify the SDE greater than 90% at 2.1 K, we further examined the SDE of device 02#F9 cooled using a G–M cryocooler (base temperature of 2.13 K), which gave a value of 90.1% at a DCR of 10 Hz (see Supplementary Fig. 6). To the best of our knowledge, this study is the first report of an NbN SNSPD with an SDE greater than 90%. More importantly, it was achieved using a G–M cryocooler (approximately 2.1 K) and therefore represents a practical and affordable potential solution for single-photon detection for quantum information.

The DCR increased monotonously with $I_b$ at all temperatures. Two different regions were clearly distinguished: background blackbody-radiation-dominated DCR (low-bias zone) and intrinsic vortex-related DCR (high-bias zone). The intrinsic DCR logarithmically increased with increasing $I_b$, which shifted to the high bias current with an increase/decrease of $I_{sw}$/temperature. Its onset $I_b$ depends on the temperature. For example, in the case of device 02#F9 (04#E4) the onset $I_b$ was suppressed to 0.95(0.97)$I_{sw}$ at the lowest temperature of 16 mK; by contrast, it was 0.89(0.92)$I_{sw}$ at 3 K. The background DCR curves followed the same curve; however, the highest background DCR slightly increased with increasing/decreasing $I_{sw}$/temperature. The same background DCR value at the bias current can be explained by the consistent SDE at the same bias current and stable background blackbody radiation.

Figure 2(c) shows the temperature dependence of the SDEs for devices 02#F9 and 04#E4 at a DCR of 10 Hz. The green-dashed line indicates the lowest temperature that a commercial G–M cryocooler (e.g., SHI RDK-101D from Sumitomo Inc.[16, 19]) can reach directly. For both devices, the SDE increased with decreasing temperature and saturated at low temperatures. The onset temperature ($T_{os}$) where the SDE reached saturation with decreasing temperature is indicated by the arrow in Fig. 2(c). The $T_{os}$ of 02#F9 and 04#E4 were 1.8 K and 1.3 K, respectively. Because of the wide saturation plateau, the SDE($T$) of 02#F9 exhibited a slow decrease with increasing temperature, in contrast to that of 04#E4. We attributed this different saturation behavior to differences in film thickness, which will be discussed later.

Figure 2(d) shows the temperature dependence of $I_{sw}$s for devices 02#F9 and 04#E4. The measurement error of the $I_{sw}$ values was approximately ±0.05 μA. When the temperature was lower than the $T_{os}$, the $I_{sw}$ gradually reached a maximum. The $I_{sw}$s at $T_{os}$ were 14.5 and 21.9 μA, respectively. A fitting based on the Bardeen model[20] simplified as $I_{cd} = \alpha \cdot (1 - (T/T_c)^2)^{3/2}$ was applied to the measured $I_{sw}$ values; the fitting results obtained with $\alpha$ and $T_c$ treated as variable parameters are plotted as solid lines in Fig. 2(d). These fitting results are consistent with the measured data, except for a slight deviation in the low-temperature region ($T < 1$ K). Given the possibility of constrictions in the nanowire, which may limit the experimental maximal $I_{sw}$, we speculate that the measured $I_{sw}$ could be smaller than the calculated depairing current $I_{cd}$.

To experimentally study the performance of the optical design, we measured the wavelength dependence of the maximum $SDE_{//}$ and $SDE_{\perp}$ when the polarization of the photons was parallel and perpendicular, respectively, to the direction of the nanowire. For comparison, the simulated absorption $A$ is also given for the wavelength range from 1200 to 1800 nm. As shown in Fig. 3, the absorption $A_{//}$ showed a broad peak with a value of 99.5% (98.1%) for 7(8)-nm-thick nanowire at an incident photon wavelength of approximately 1550 nm. The measured $SDE_{//}$s were approximately 7% lower than the calculated absorption values, whereas the $SDE_{\perp}$s were approximately 6% higher than the corresponding calculated absorption values. That is, the measured polarization extinction ratio (PER, defined as $SDE_{//}/SDE_{\perp}$) was lower than the calculated PER (defined as $A_{//}/A_{\perp}$). For example, in the case of 7(8)-nm-thick devices, the experimental PER of 3.5(3.2) at 1550 nm was lower than the calculated PER of 4.4(4.6). This deviation indicates that the practical device differs from the design in geometry and/or material optical parameters. Further optimization of the fabrication process will be helpful in further improving the consistency and the SDE. In addition, an extra ARC upon the nanowire can suppress the PER by a factor of two or more, if necessary (Supplementary Table 1). However, the SDE decreased faster than the simulated values when the wavelengths were longer than 1600 nm, which can be explained by a non-unity $\eta_{int}$ due to the smaller photon energy (Supplementary Fig. 8).

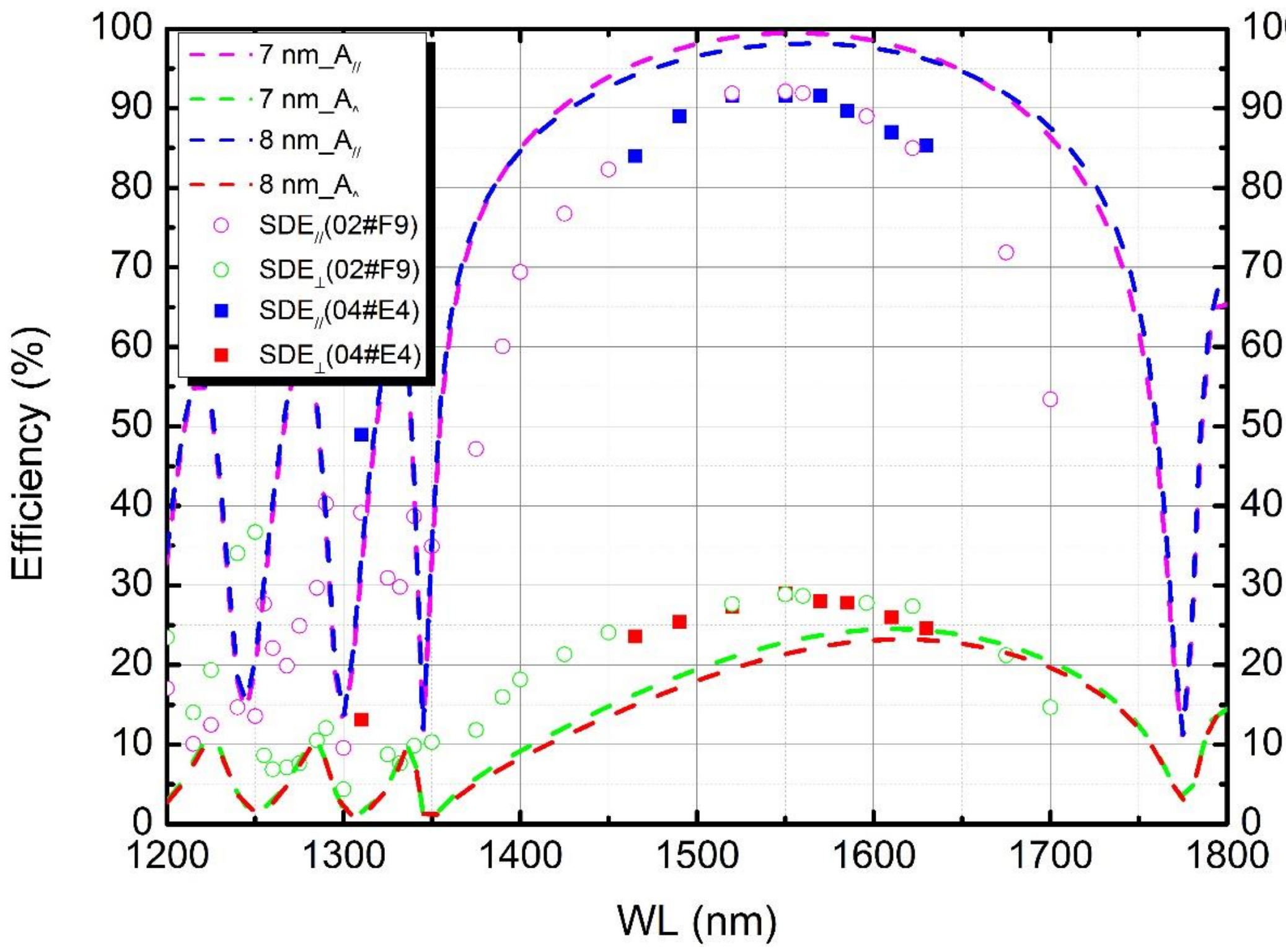

Fig. 3. (Color online) Wavelength dependence of the measured SDEs for devices 02#F9 (open circles) and 04#E4 (solid squares) in the wavelength range from 1200 to 1800 nm. The SDEs were recorded when the bias current was 14.0 and 21.0 μA, respectively, with a DCR of 10 Hz. The dashed lines show the simulated parallel ($A_{//}$, magenta and blue) and perpendicular ($A_{\perp}$, green and red) polarized absorption values as functions of wavelength for 7- and 8-nm-thick nanowires, both with a 75-nm width and a 140-nm pitch.

The aforementioned results indicate that the NbN SNSPD exhibited an SDE of 90% at the dark count rate of 10 Hz at 2.1 K. The other important parameter, timing jitter $T_j$, was also measured at 16 mK and 2.1 K. Figure 4 shows the $T_j$ of the SNSPDs (device 02#F9 and 04#E4), as measured at a bias current of 14.5 and 21.5 μA (both at $0.95I_{sw}$), respectively, using a time-correlated single-photon counting (TCSPC) module (SPC-150, Becker & Hickl GmbH). A 1550-nm-wavelength pulsed laser with a 100-fs pulse width was used, and the input photon flux was attenuated to the single-photon level. For devices 02#F9 and 04#E4, the $T_j$ values defined by the full-width at half-maximum (FWHM) of the histogram were 70.2 and 40.0 ps at 16 mK and increased to 79.0 and 46.0 ps at 2.1 K, respectively, because of the reduction of $I_b$. Notably, the values of $T_j$ measured using the CDR were a few picoseconds larger than those measured using the G–M cryocooler at the same bias current because of the long coaxial cable (about 4.5 m) used with the cryogenic system. Nonetheless, the measured $T_j$ of the NbN SNSPD was approximately two times smaller than that of the WSi SNSPD because of its much larger switching current (by a factor of approximately three). Figure 4(b) shows the bias-current dependence of $T_j$; this figure implies that the signal-to-noise ratio plays the key role in determining $T_j$. A comparison of more parameters among the Nb(Ti)N-, WSi-, MoSi-SNSPDs, and W-TES is provided in Supplementary Table 2; this comparison indicates that our NbN detectors represent a substantial advancement.

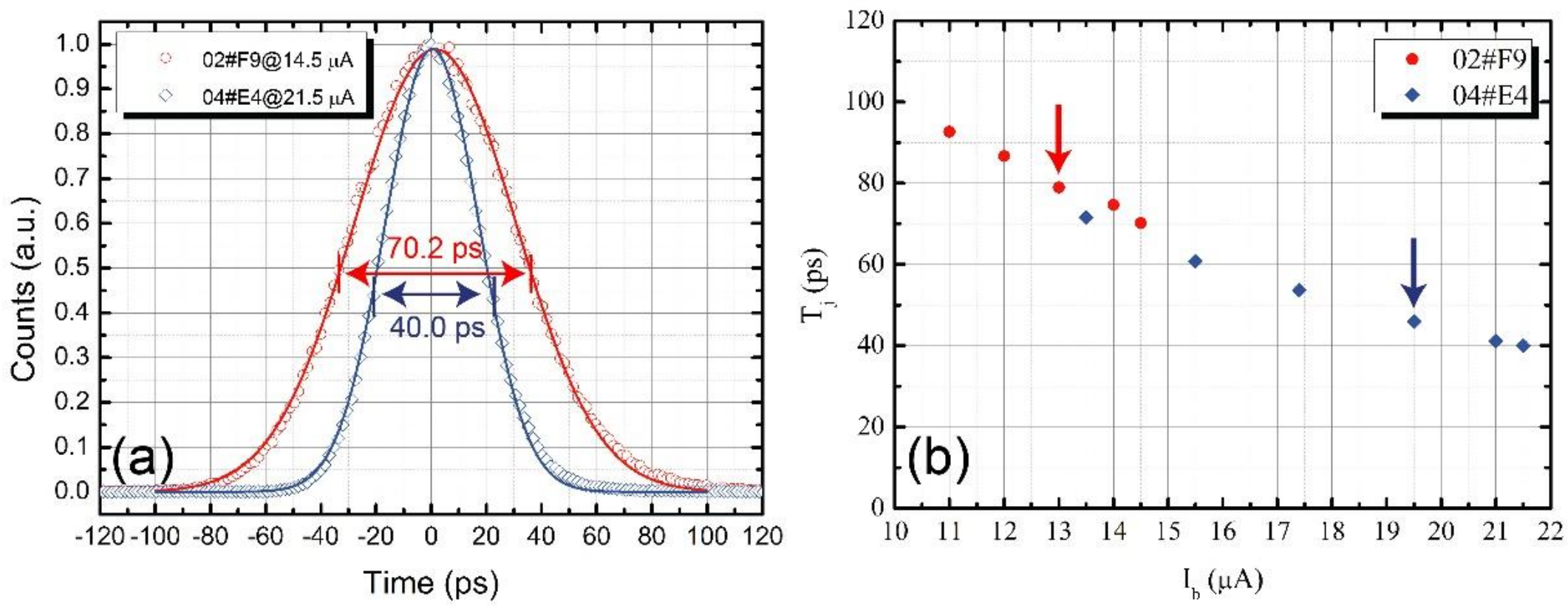

Fig. 4. (Color online) (a) Timing jitter ($T_j$) measurement of two devices measured at 16 mK. For devices 02#F9 and 04#E4, the corresponding FWHM values of the Gaussian fits of $T_j$ are 70.2 and 40.0 ps, respectively (see solid lines). (b) Bias current dependence of $T_j$. The $T_j$ increased to 79.0 and 46.0 ps at 2.1 K (indicated by arrows) because of the reduction of $I_b$ to 13.0 and 19.5 μA for devices 02#F9 and 04#E4, respectively.

**Discussion**

Four of the approaches used in this study contributed to the remarkable improvement of the SDE. The first approach was the DBR cavity design, which ensured near-unity reflectance of 1550-nm photons. The second approach was the removal of the reflection of the substrate in the backside fiber-coupled SNSPDs via front-side lens fiber alignment. Third, the use of lens fiber guaranteed better optical coupling between the nanowire and the incident photons because of a smaller beam size than that achieved with an SM-28e fiber. The fourth approach was to maximize the optical absorption and IDE simultaneously by tuning the thickness while leaving the nanowire linewidth fixed. Optimization was achieved with a nominal film thickness of 7 nm. The absorption reached a maximum, and the IDE approached unity even at 2.1 K. In the case of the 8-nm-thick film, less saturated SDE behavior was observed at 2.1 K, though the maximum SDE was 91.7% at 16 mK. With the thinner film thickness of 6.5 nm, a more saturated SDE behavior was registered for SNSPD at 2.1 K; however, the highest SDE was only 82%, which indicates low optical absorption (Supplementary Fig. 7).

In conclusion, we successfully demonstrated that the SDE of an NbN SNSPD surpassed 90(92)% at a DCR of 10 Hz at 2.1(1.8) K because of the adoption of a DBR mirror and the simultaneous maximization of the optical absorption and IDE. In addition, the NbN SNSPD exhibited a timing jitter of only 79 ps. We believe some room remains for further improving the SDE to near unity and achieving a higher operating temperature through optimization of the fabrication process and the optical and superconducting properties of the film.

## Methods

**Device preparation.** A series of NbN films with various nominal thicknesses from 6.5 nm to 8 nm were deposited onto 2-inch DBR wafers at room temperature. NbN thin films were deposited in a reactive direct-current magnetron sputtering system in which the partial pressures of Ar and $N_2$ in the Ar–$N_2$ gas mixture were 79% and 21%, respectively. The thickness of the films was controlled by the sputtering time and was verified by XRD analysis. The NbN films were then patterned into a meandered nanowire structure using electron-beam lithography (EBL) with a positive-tone ZEP520A electron-beam resist and were reactively etched in $CF_4$ plasma. The width of the nanowires was fabricated to approximately 75 nm with different pitches of 130, 140, 160, 180, 200, and 230 nm. Finally, a 50-Ω-matched coplanar waveguide was formed using ultraviolet lithography and reactive-ion etching. The active areas of the SNSPDs were designed with two sizes (⌀15 and ⌀18 μm), both of which are much larger than the 6.8-μm mode field diameter of a lens fiber for an incident photon wavelength of 1550 nm. The lens fiber was designed with a focal distance of approximately 145 μm, an ARC target wavelength of 1550 nm, and reflectance less than 1%. The fiber was aligned to the center of the active area using a microscope at room temperature, and the alignment error was estimated to be ±3 μm. The devices with the lens fiber were packaged in a copper box.

**Measurement setup.** Figure 6 shows a schematic of our measurement system incorporating a CDR. The fiber-coupled package was mounted onto an Au-plated oxygen-free Cu platform thermally connected to the mixing chamber of the CDR with a base temperature of 10 mK. The sample temperature was indicated by a $RuO_x$ resistance thermometer positioned at the bottom of the mixing chamber. The dilution unit was inside a refrigerated dewar mounted onto a vibration-attenuated stage. The temperature was tuned using a proportional–integral–derivative (PID) controller. The temperature stability varied by region: $T \leq 1$ K, ±0.001 K; $T \leq 1.8$ K, ±0.04 K; $T \approx 2$ K, ±0.06 K; $T \geq 2.3$ K, ±0.03 K.

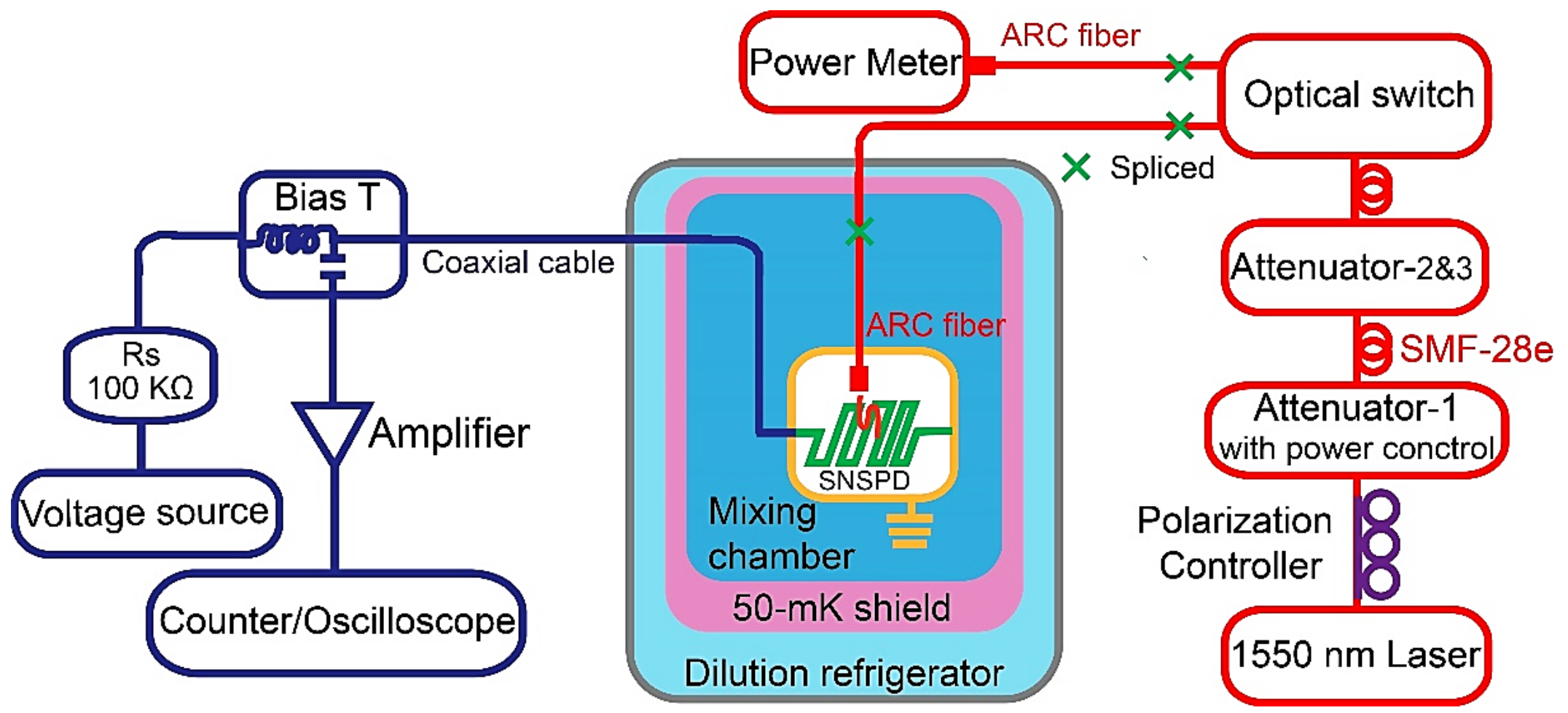

Fig.5. (Color online) Schematic of the ultralow-temperature measurement system used to characterize the SNSPDs. The SNSPDs are mounted on the plate of the mixing chamber; the plate can easily control the temperature from 16 mK to 8 K via the PID controller. Optical (represented by red lines) and electrical (represented by blue lines) components are indicated in the figure.

A spliced single-mode (SM) optical fiber and a 50-Ω coaxial cable were connected from the top of the fridge to the SNSPD package. The coupling loss due to the fiber splicing was typically approximately 0.01 dB. To suppress the blackbody radiation of the fiber, the SM fiber was coiled to a diameter of approximately 30 mm in the mixing chamber. The optical loss due to the coils was carefully checked and was observed to be less than 0.07 dB at an incident photon wavelength of 1550 nm. Two tunable-wavelength continuous-wave (CW) laser diodes (Keysight 81970, 1465–1575 nm; Keysight 81940, 1520–1630 nm) and a supercontinuum white-light laser (NKT photonics, EXB-3) were used as the photon sources. The input light was heavily attenuated by three cascaded attenuators so that the photon flux at the input connector of the cryostat was single-photon level. A polarization controller (Agilent N7786B) was inserted in front of the attenuator, together with power controller, to obtain the maximum SDE. Each time the polarization was tuned, the input power was carefully rechecked by switching the input photons to the power meter (Keysight 81624B) port. The split ratio of the MEMS optical switch was 1:1.002 ± 0.002, which guaranteed nearly identical numbers of photons input to the power meter and the SNSPD. The power calibration was performed at a level of −38.930 ± 0.006 dBm, which was within the linear region of the power meter. A cascaded attenuation of −33 and −40 dB to the input photons was then applied using two attenuators, and the final input power reached −111.930 dBm, corresponding to a photon flux of $5 \times 10^4$ photons/s. The linearities of the three attenuators were recalibrated using the power meter; they were found to be 0.999 ± 0.001 (see Supplementary Fig. 5).

The device was current-biased through the dc arm of the bias tee. The response pulse of the detector was readout via the ac arm of the bias tee using a room-temperature amplifier (RF Bay Inc., LNA650) and photon counter (SRS Inc., SR400). The SDE was defined as SDE = (PCR − DCR)/($5 \times 10^4$), where PCR is the output pulse rate of the SNSPD.

The SDEs included all losses in the system, and the overall relative errors of the SDE values were estimated to be 2.82%, resulting from the relative uncertainty of the power meter calibration (2.80%)[21], the uncertainty of the splitting ratio (0.20%), uncertainty of the attenuation (0.33%), and the uncertainty of the PCR (0.10%). The dominant uncertainty was from the light power calibration; this uncertainty could be further reduced using the Agilent calibration option C05[21] or the correlated-photon-based method[22].

**Acknowledgments**

The authors would like to thank M. Wang, Z. J. Li, and B. Gao from SIMIT for technical support during the ultra-low temperature measurements. This study was funded by the National Natural Science Foundation of China (91121022, 61401441, & 61401443), Strategic Priority Research Program (B) of the Chinese Academy of Sciences (XDB04010200), and the Science and Technology Commission of Shanghai Municipality under Grant 16JC1400402.

**Author contributions**

W.J.Z. designed and fabricated the devices. W.J.Z. and L.X.Y. wrote the paper. L.X.Y. organized the research. H.L. preformed numerical simulations. W.J.Z. and J.H. performed the measurements. C.L.L helped to perform the ultra-low temperature measurements with constructive discussions. L.Z. and J.H. provided the NbN films. X.Y.L, J.J.W, Z.W., X.M.X., and M.H.J provided helpful discussions. All authors reviewed the manuscript.

**Additional information**

Competing financial interests: The authors declare no competing financial interests.

**Supplementary Information**

**Surface morphology of the NbN films**

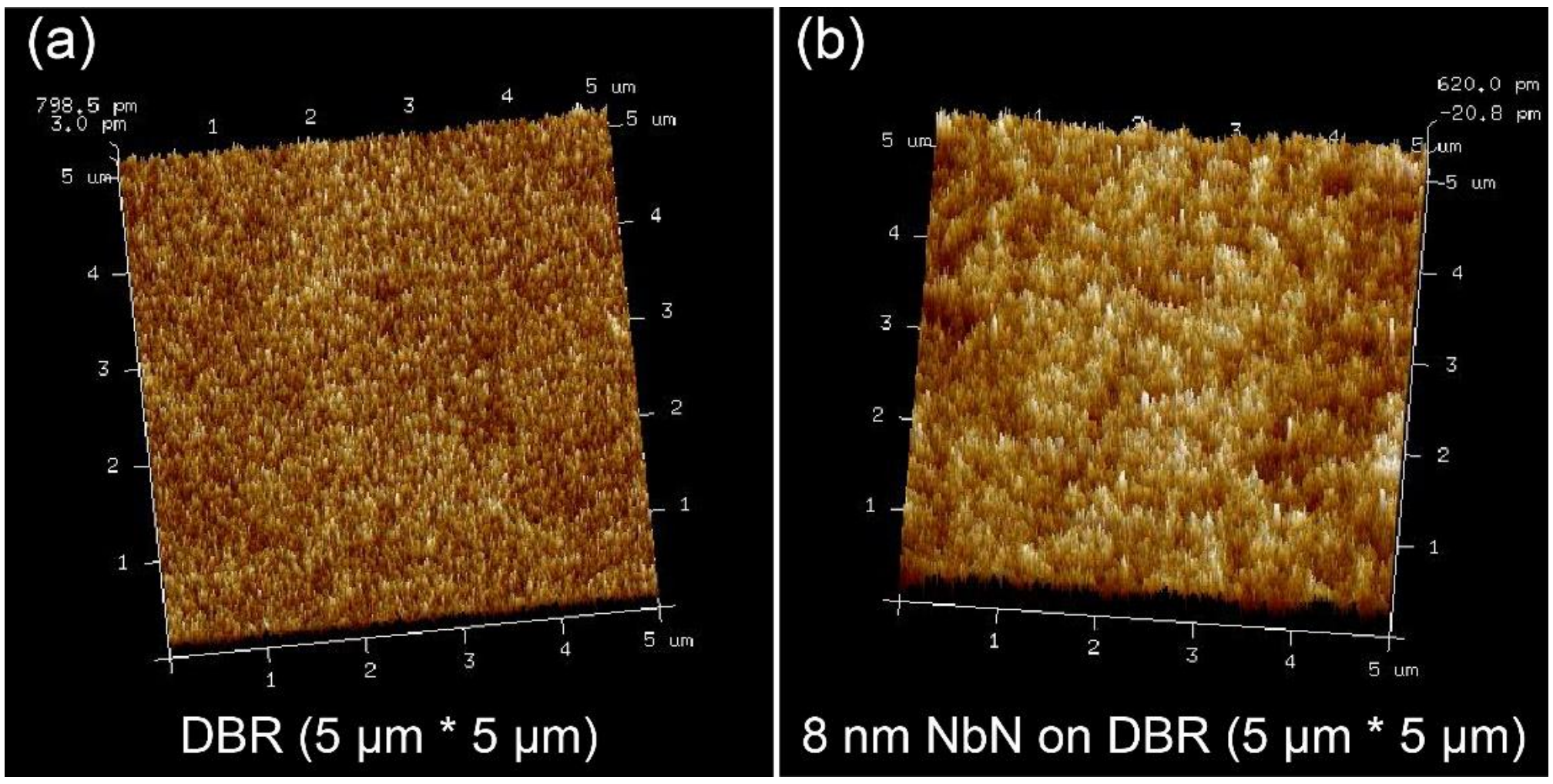

Supplementary Figure 1 (Color online) Atomic force microscopy (AFM) images of the DBR substrate and the 8.0-nm NbN film deposited onto DBR. The roughness of the DBR substrate and NbN film were approximately 0.2 and 0.3 nm, respectively.

To characterize the uniformity of the film thickness, we conducted high-resolution atomic force microscopy (AFM) measurements. The surface morphology of the DBR substrate and an 8.0-nm NbN thin film deposited onto DBR are presented in Supplementary Fig. 1. From the AFM image, RMS (root mean square) surface roughness of 0.2 and 0.3 nm were determined. The roughness of the NbN film was slightly larger than that of the DBR because of the lattice mismatch between them. The measurements revealed that thickness variation of the NbN film was less than a few percent.

**Measurement of reflectance of NbN films on DBR substrates**

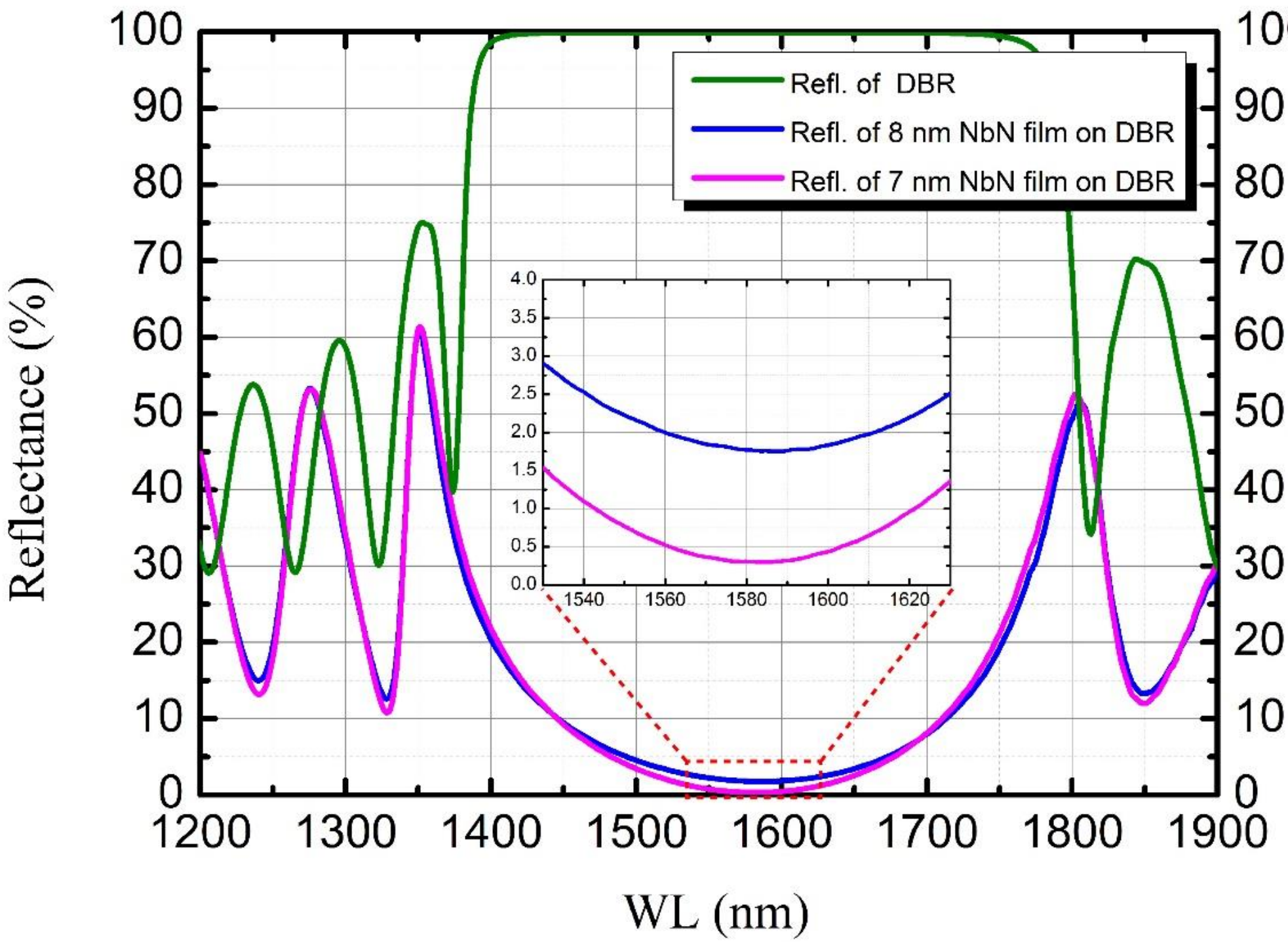

Supplementary Fig. 2. (Color online) Wavelength dependence of the reflectance of a DBR wafer ($R_{DBR}$) and 7- and 8-nm-thick NbN films deposited onto DBR wafers ($R_{NbN}$), as measured at room temperature by a spectrophotometer. At an incident photon wavelength 1550 nm, the $R_{DBR}$ and $R_{NbN}$ for 7(8) nm were 99.9% and 0.77(2.24)%, respectively. The inset shows a zoomed-in region around the $R_{NbN}$ minima of the 7(8)-nm films.

Because the absorption $A$ can be expressed as $A = 1 - Tr - R$ and the transmission $Tr$ is negligible (approximately 0.1%) for the high-reflectance mirror, the reflectance $R$ can be directly measured using a spectrophotometer; $A$ can then be deduced as $A \approx 1 - R$. Figure 4 shows the measured reflectance as a function of wavelengths from 1200 to 1900 nm for the with 7(8)-nm-thick NbN films deposited onto DBR mirrors. The reflectance minima, which appear at approximately 1570–1600 nm, imply high absorption of the NbN films in this wavelength range.

The measured reflectance data show that near-unity absorption in NbN films can be attained when the film thickness on the DBR mirror is approximately 7 nm. With increasing thickness, the reflectance increases. Notably, however, the absorption of a thin film ($A_{film}$) can differ from the absorption of a nanowire or device ($A_{device}$), and the $A_{film}$ defines the upper bound for $A_{device}$.

**Simulation of absorption for devices**

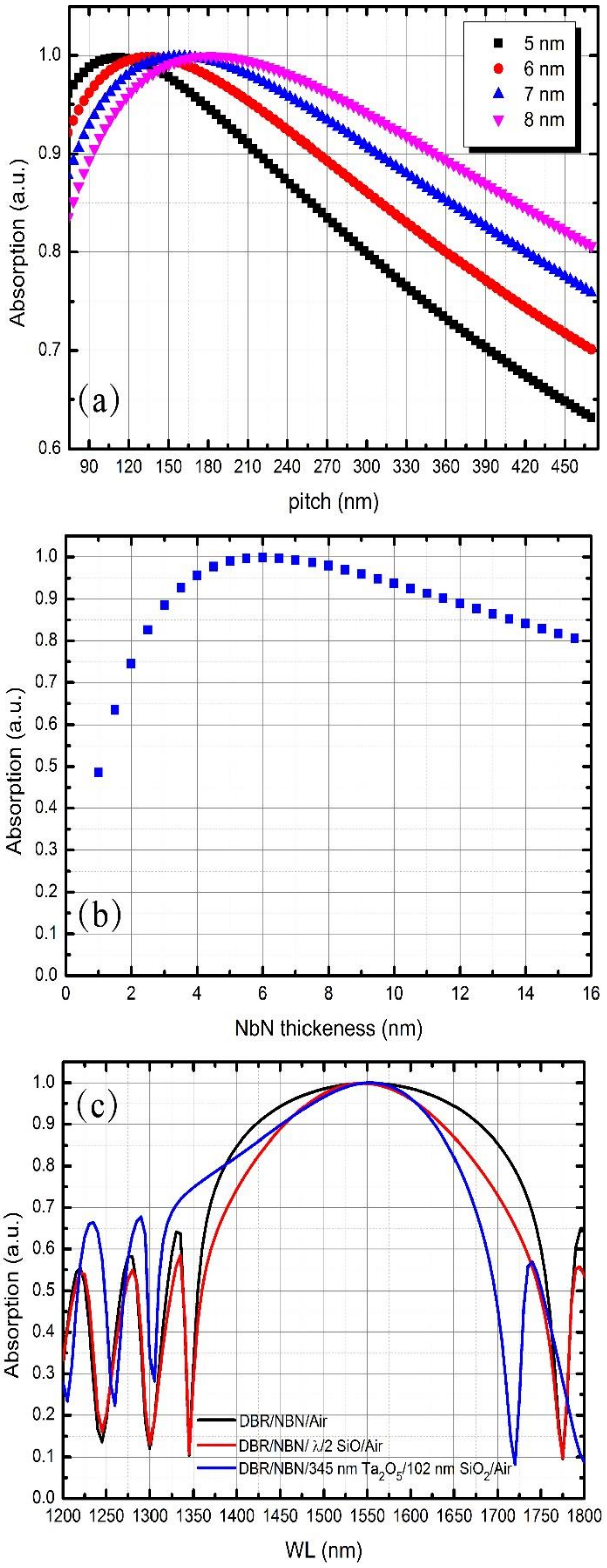

Supplementary Figure 3. (Color online) (a) Simulated absorptions of devices as a function of the pitch with a fixed width (75 nm) and thickness (5, 6, 7, 8 nm). (b) Simulated absorptions of devices as a function of the NbN thickness, with a fixed width (75 nm) and pitch (140 nm). (c) Wavelength dependence of absorptions for the half-cavity (black line)

and full-cavity (red, blue, and green lines) devices, with a fixed width (75 nm) and pitch (140 nm). For the full-cavity designs, the front anti-reflection coating (ARC) was tuned through the use of various dielectric materials. The ARC introduced to the half-cavity design resulted in narrowing of the bandwidth.

To optimize the $A_{device}$, we performed numerical simulations using the rigorous coupled-wave analysis method. The parameters in the simulation; i.e., the refractive index of NbN, $n_{NbN}$ = 6.50 + 5.83$i$, at 1550 nm was obtained using a spectroscopic ellipsometer, and the refractive index and thickness of the Si substrate were $n_{Si}$ = 3.46 and $L$ = 400 μm, respectively.

Supplementary Fig. 3(a) shows the $A_{device}$ as a function of pitch with a fixed 75-nm width for the case at different thicknesses (5.0–8.0 nm). In the case of a thickness of 7.0 nm, maxima greater than 99% appeared at pitches ranging from 130 to 190 nm. As the thickness increased, the position of the maxima shifted to a larger pitch; i.e., a larger spacing was necessary for thicker films. As shown in Supplementary Fig. 3(b), the $A_{device}$ with fixed width and pitch (75 and 140 nm) first increased then fell gradually as the NbN thickness increased. $A_{device}$ reached a maximum greater than 0.99 at 5.5–7.0 nm.

Supplementary Fig. 3(c) shows the influence of the antireflection coating (ARC) on the half-cavity (DBR/NbN/Air) design. The ARC designs were tuned through the use of various materials, including SiO (the SiO refractive index $n_{SiO}$ = 1.89), $SiO_2$, and $Ta_2O_5$. We observed that the ARC coatings (SiO and $SiO_2/Ta_2O_5$ bilayer) did not substantially change the absorption around the target wavelengths but resulted in a reduction of the bandwidth. Therefore, for a simplified fabrication process, the half-cavity design can provide sufficient absorption (greater than 0.995).

**RT measurements of the devices with various thicknesses**

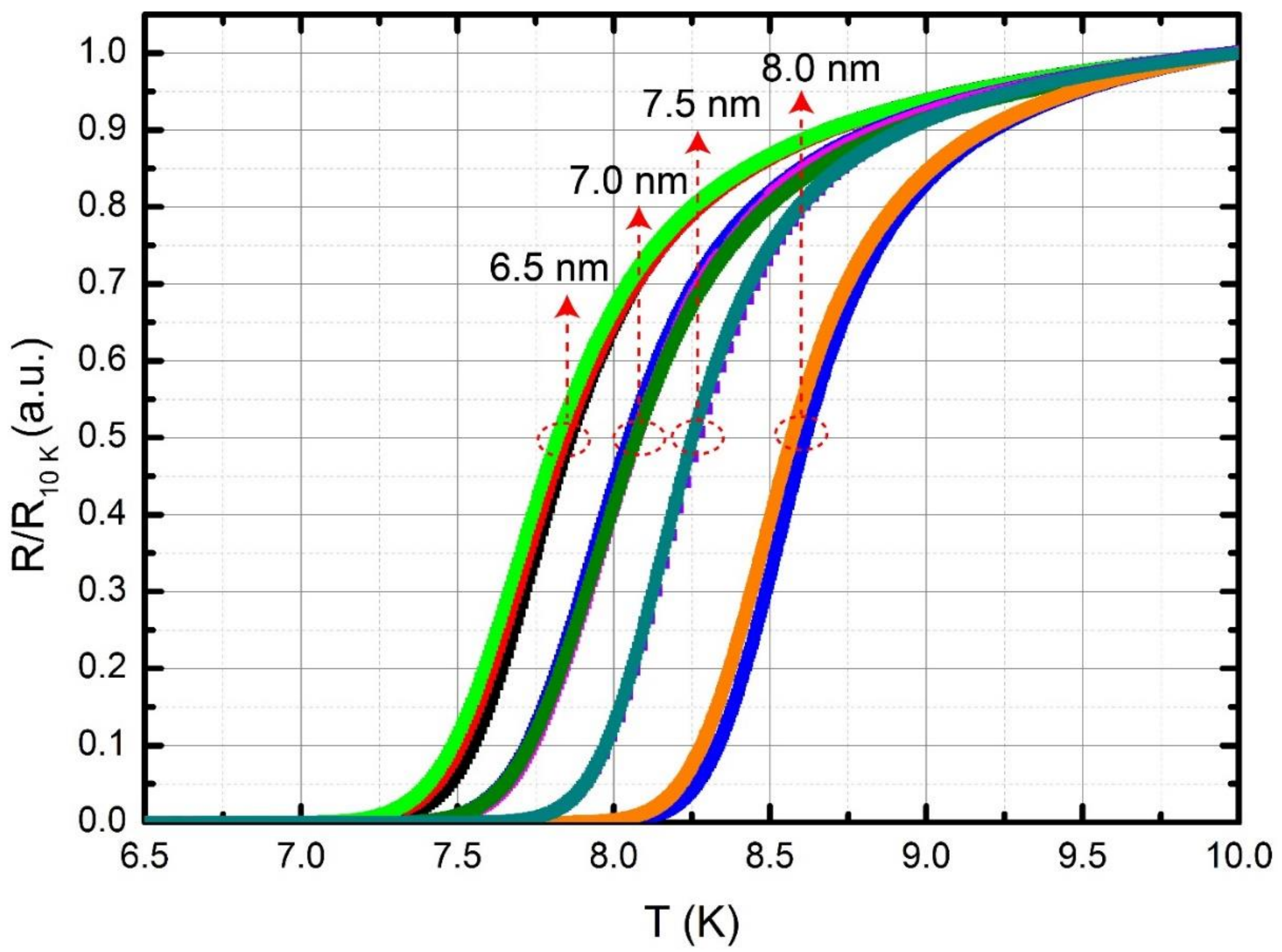

Supplementary Figure 4. (Color online) Normalized resistances of 10 devices with NbN thicknesses of 6.5, 7.0, 7.5, and 8.0 nm as functions of temperature. The resistance $R$ was normalized to the resistance value at 10 K ($R_{10K}$). The $T_c$s (defined by $0.5R/R_{10K}$) of 6.5, 7.0, 7.5, and 8 nm devices were 7.85, 8.07, 8.26, and 8.59 K, respectively, with an error of approximately ±0.05 K.

Supplementary Fig. 4 shows the RT measurements for 10 devices with different nominal thicknesses from 6.5 to 8.0 nm. The curves were normalized by their resistance values at 10 K ($R_{10K}$). The $T_c$s of the thin films were determined by the criterion $0.5R/R_{10K}$. The deposition rate of the NbN films was approximately 0.8 nm/s, and the thickness was controlled via the deposition time. Because these NbN films were fabricated in the same run, clear separation of the RT curves for each thickness was obtained, indicating good control of the deposited NbN thickness.

**Calibration of the optical attenuators**

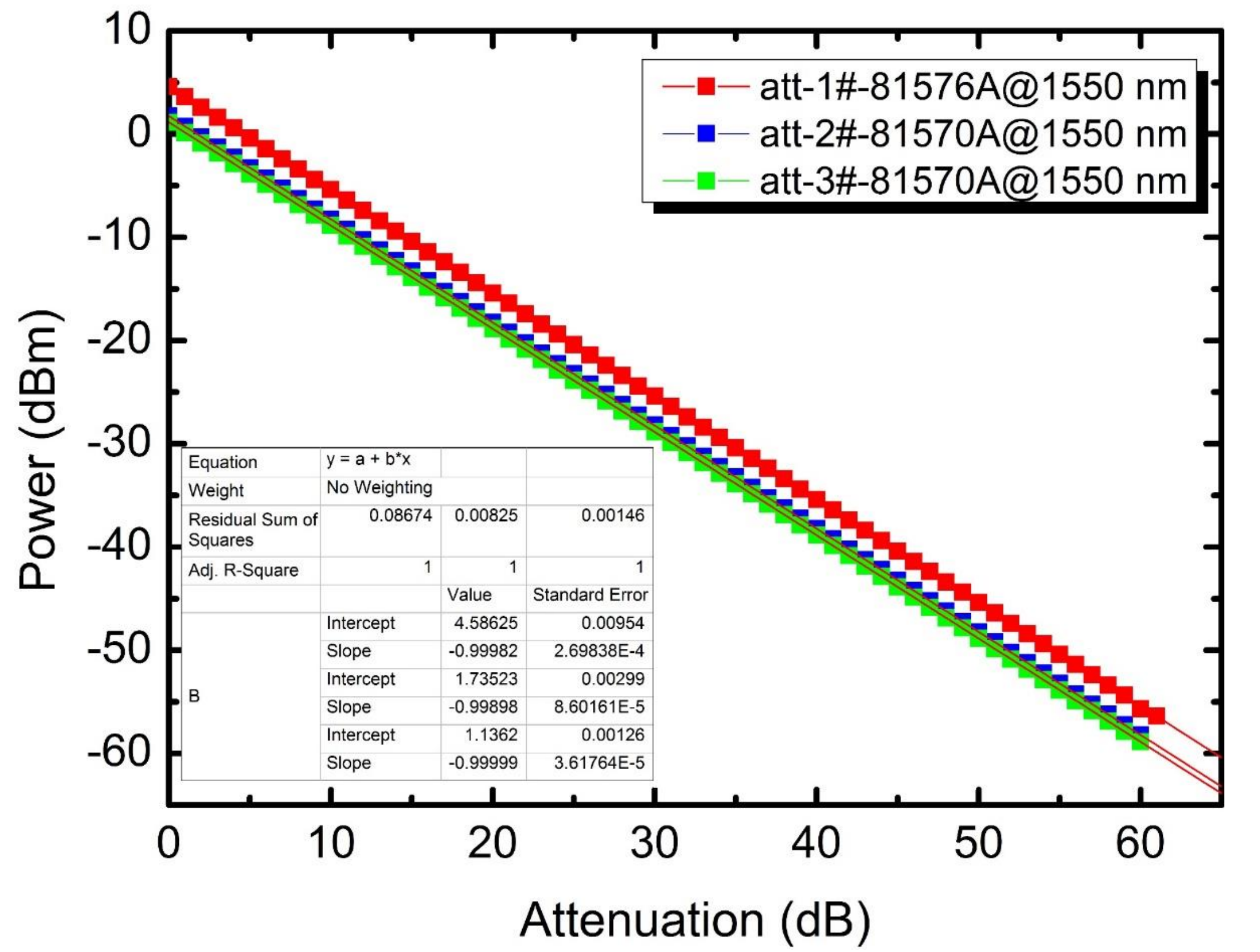

Supplementary Figure 5. (Color online) Measurement of the individual attenuator linearity at a wavelength of 1550 nm as the input power as varied. Three independent attenuators supplied as much as 180 dB of optical power reduction. All of the output powers were measured using the same power meter (Keysight 81624B). All the attenuators exhibited good linearity that approached unity, independent of the input optical power.

Three attenuators (one 81576A, two 81570As) were calibrated separately to confirm the expected linearity, particularly at high attenuation values. Supplementary Fig. 5 shows the measured transmitted power as a function of the nominal attenuation. The attenuation was measured at input power levels of 1–5 dBm, which emitted from the tunable laser source (Keysight 81940A) employed in the SNSPD characterizations. The linear fitting revealed that the linearity of the attenuators was at the level of $0.999 \pm 0.001$, ensuring precise attenuation.

**SDE and DCR performance of SNSPD 02#F9 cooled by a G–M cryocooler**

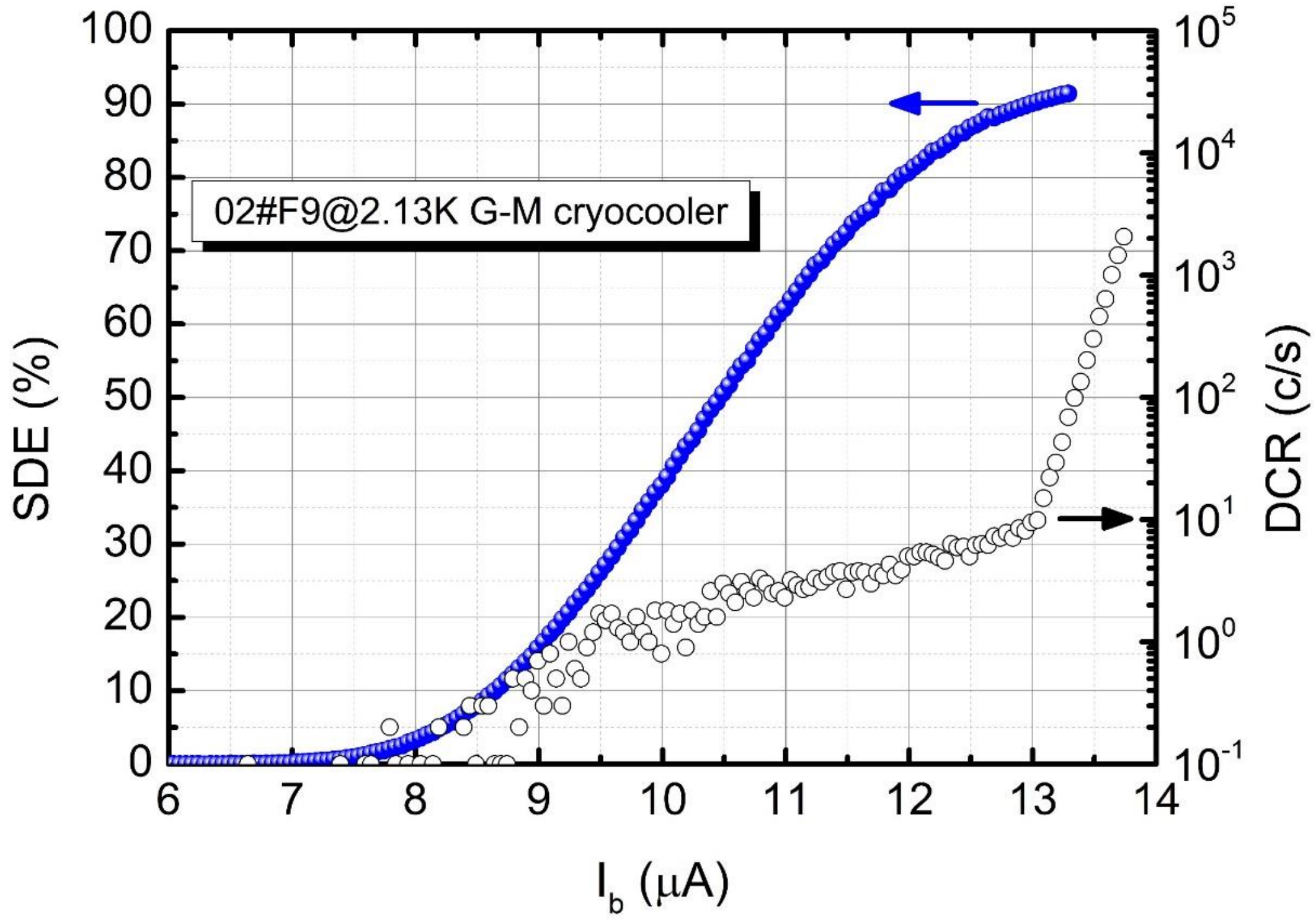

Supplementary Figure 6. SDE and DCR vs. $I_b$ for device 02#F9 measured at 2.13 K using a G–M cryocooler. The SDE of 90.1% was obtained at a DCR of 10 Hz.

After the ultra-low temperature measurements, the device 02#F9 was mounted in a 2.125 ± 0.005 K G–M cryocooler. Supplementary Fig. 6 demonstrates the SDE and DCR as functions of $I_b$. The measurement results are consistent with those measured in CDR operating at 2.10 ± 0.06 K. The $I_{sw}$s obtained using the CDR and G–M were 13.8 and 13.7 μA, respectively, with a standard error of approximately 0.05 μA. The SDE of 90.1% was obtained at a DCR of 10 Hz using a G–M cryocooler.

**Performance of 6.5-nm-thick devices**

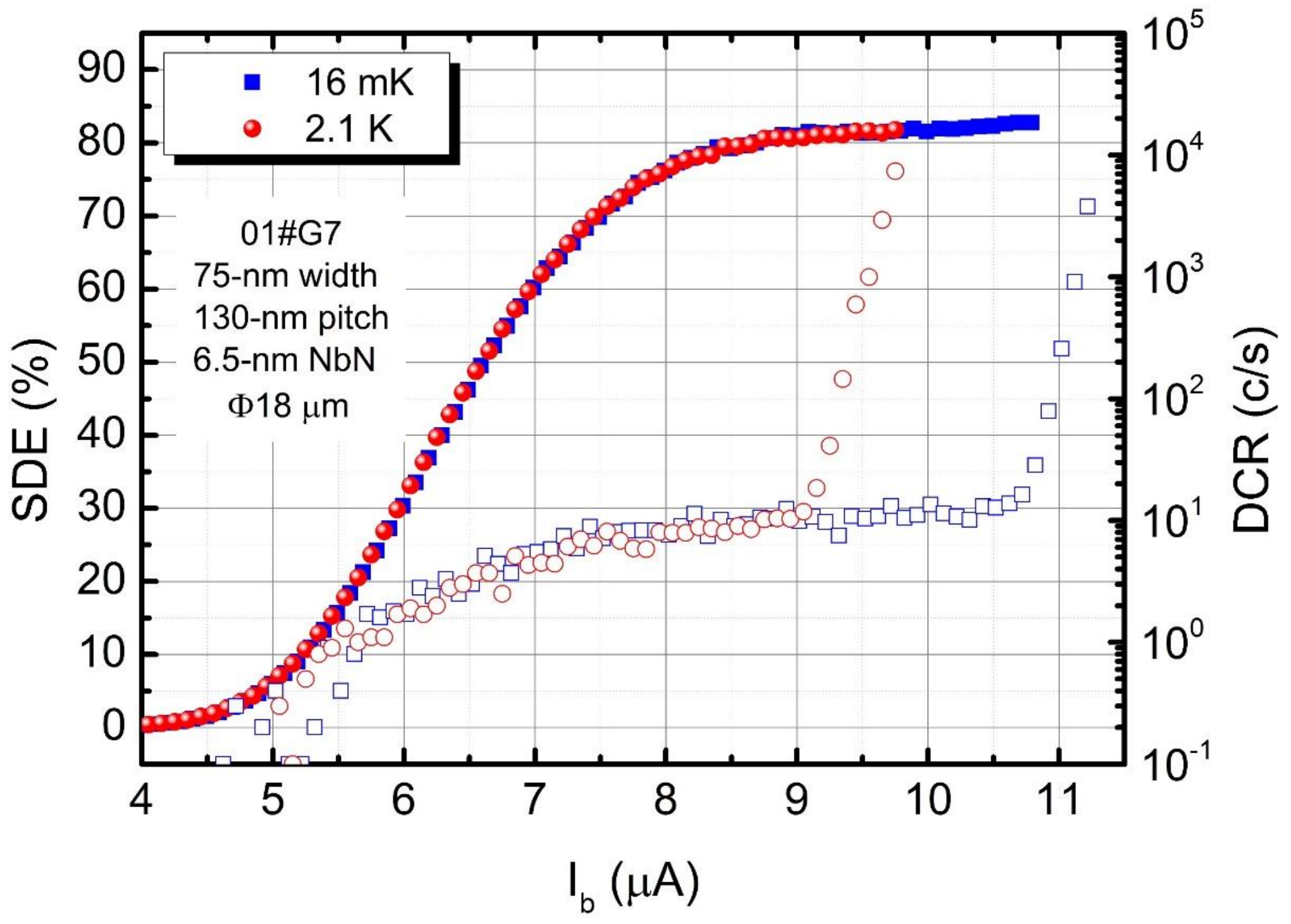

Supplementary Figure 7. (Color online) Parallel polarized SDE and DCR vs. $I_b$ for device 01#G7 with an NbN thickness of 6.5 nm, as measured at 2.1 and 16 mK. Because of the reduction of thickness, a clear saturation plateau is still observed at 2.1 K.

Supplementary Fig. 7 shows the parallel polarized SDE and DCR as functions of $I_b$ for device 01#G7 measured at 2.1 and 16 mK. This device features a nominal 6.5-nm-thick, 75-nm-wide, and 130-nm-pitch nanowire covering an active area of ⌀18 μm. Because of the thinner nanowire, its $I_{sw}$ of 11.2 μA at 16 mK is lower than those described in Fig. 2 (15.2 and 22.5 μA). The maximum SDE for this device was 82.2%, and the PER was 3.9 at 1550 nm. A substantially lower SDE was obtained with the device with a 6.5-nm-thick film compared to its simulated absorption value of over 95%. This SDE was also approximately 10% lower than those of the 7(8)-nm-thick devices. We carefully measured other devices in the same run and observed saturated SDEs in the range from 70 to 82%, varying by device (Supplementary Table 1). Thus, the deviation between the experimental and simulated results might be due to the thin nanowire combined with an imperfect DBR cavity, resulting in low photon absorption.

**Wavelength dependence of SDE($I_b$)**

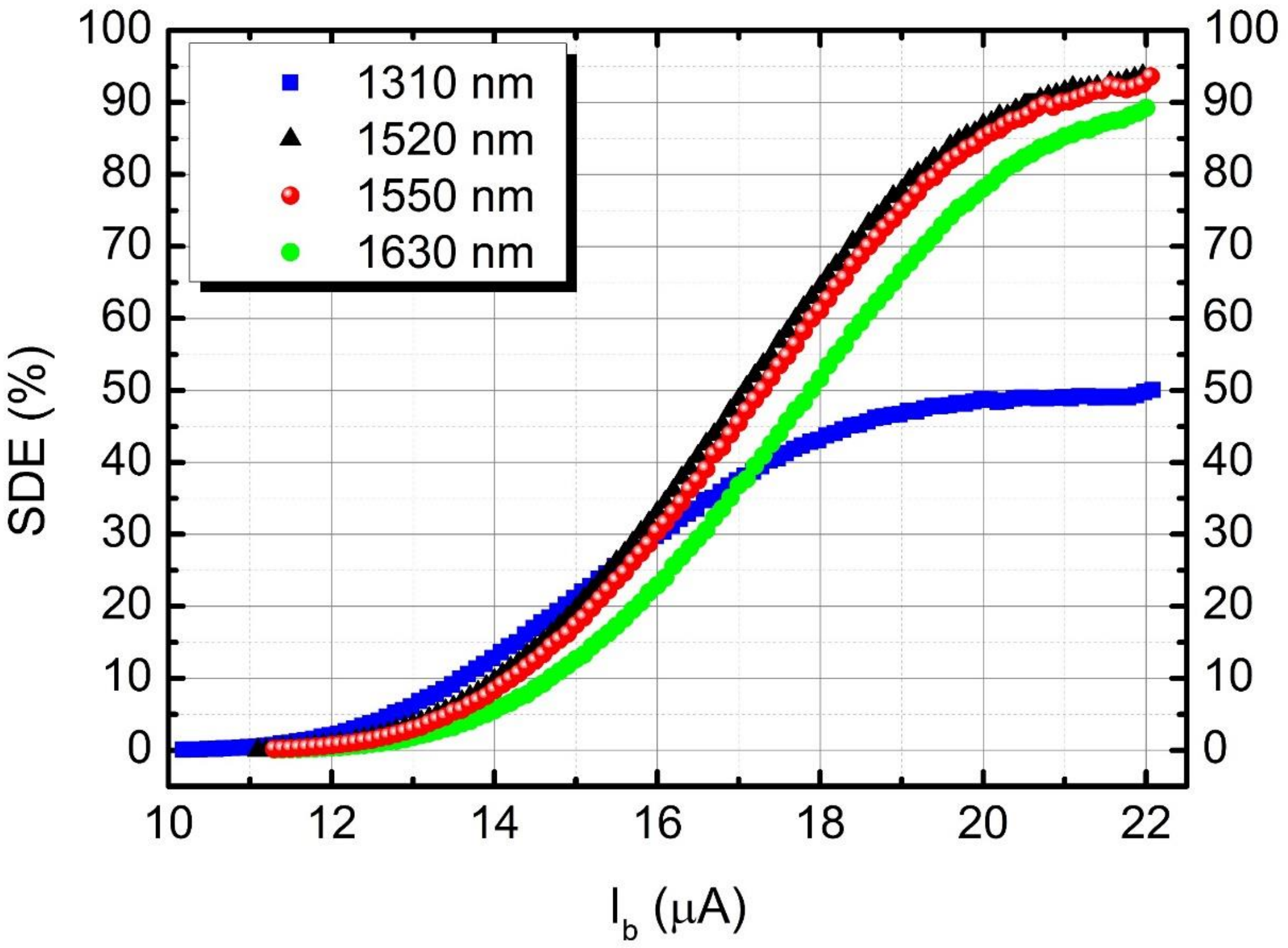

Supplementary Figure 8. (Color online) SDE vs. $I_b$ for device 04#E4, with various wavelength photons incidence, measured at 16 mK. A clear saturation plateau was obtained at 1310 nm, due to the high excitation energy of 1310-nm photons.

In order to investigate the wavelength dependence of SDE, we illuminated the device using photons of different wavelength. At each wavelength, the input photon power was carefully calibrated and attenuated to a flux of $5 \times 10^4$ photon/s. Photons with wavelengths of 1520–1630 nm and 1310 nm were emitted from a CW-laser (Keysight 81940A) and from a ps-pulse laser (Hamamatsu, C10196), respectively.

Supplementary Fig. 8 shows the SDE measured at 16 mK as a function of $I_b$ for device 04#E4, illuminated with 1310-, 1520-, 1550-, and 1630-nm photons. In the case of 1310-nm photons, a significant saturation was observed because of these photons' high excitation energy. With increasing photon wavelength, the saturation plateau shrank and eventually disappeared, indicating a non-unity IDE.

**Response pulses of photon response**

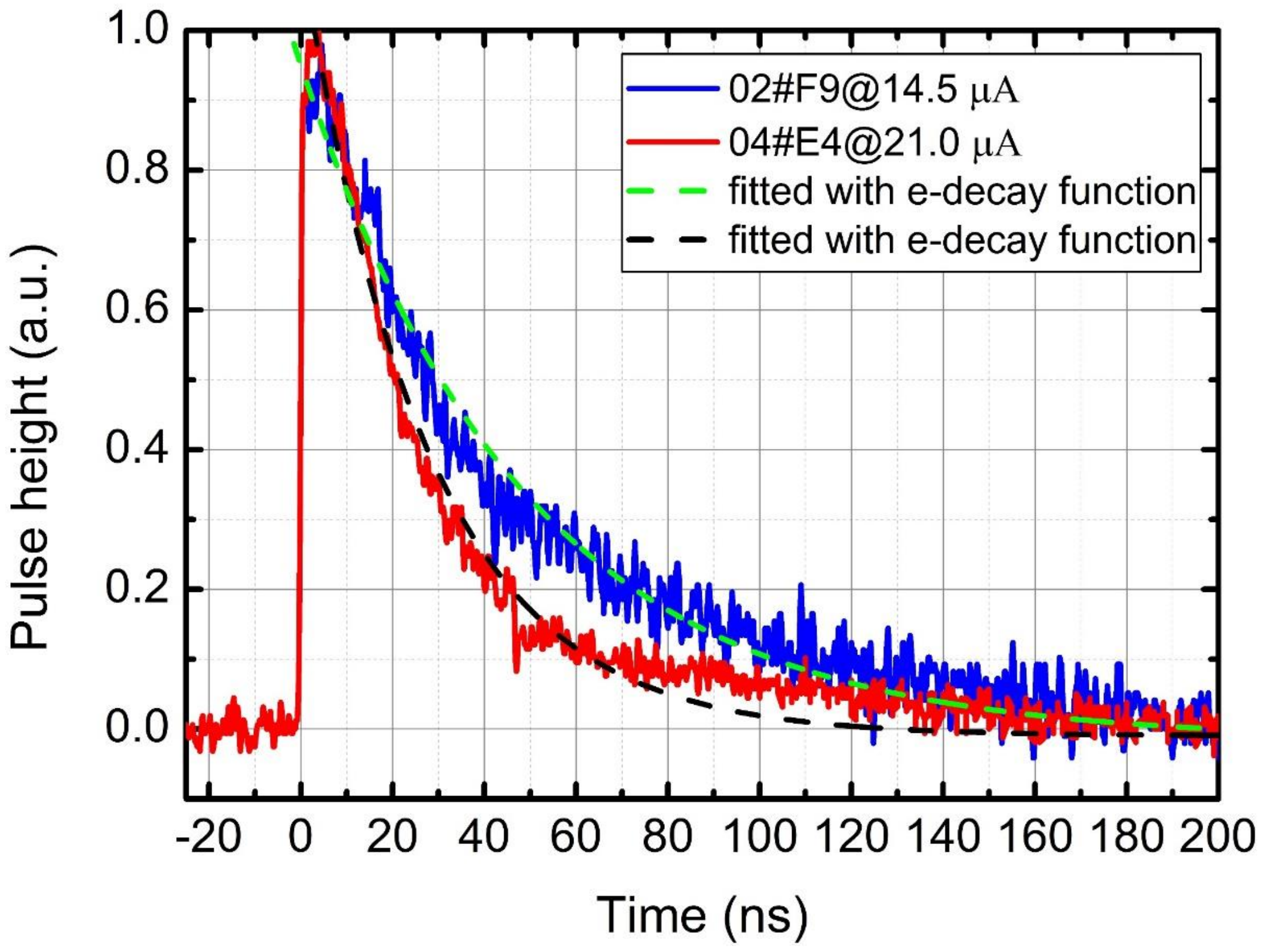

Supplementary Figure 9. (Color online) Response pulses of devices 02#F9 and 04#F9 recorded by an oscilloscope at $I_b$ = 14.5 and 21 μA, respectively, when the devices were operated at 16 mK. The fast decay time of device 04#E4 was due to its small active area of ⌀15 μm and thick thickness.

We characterized the decay time of the response pulse by directly monitoring the amplified output electronic pulse with an oscilloscope. Response pulses of two devices—02#F9 and 04#F9—are shown in Supplementary Fig. 9; the devices were biased at $I_b$ = 14.5 and 21 μA, respectively, and operated at 16 mK. We fitted the decay time using an exponential decay function expressed as $a + b(e^{-t/\tau})$, where $\tau$ is the time at which the height of the pulse is reduced to 1/e = 0.368 of its initial value and $a$ and $b$ are fitting parameters. Thus, by fitting the decay part of the pulse, we obtained the decay time $\tau$ = 48.5 (27.3) ns for device 02#F9 (04#F9).

**List of characterized devices**

| No. | Operating Temp. (K) | Thick. (nm) | W-P (nm) | Φ (μm) | $I_{sw}$ (μA) | $SDE_{max}$ (%) | PER | $SDE_{ARC}$ (%) | $PER_{ARC}$ |
|---|---|---|---|---|---|---|---|---|---|
| **1** | 0.016 | 6.5 | 75-130 | 18 | 11.2 | 82.2 | 3.9 | - | - |
| | 2.1 | | | | 10.0 | 81.4 | | | |
| **2** | 2.2 | 6.5 | 75-140 | **15** | 9.4 | 70.2 | - | - | - |
| **3** | 2.2 | 6.5 | 75-140 | 18 | 9.4 | 81.6 | - | - | - |
| **4** | 2.2 | 6.5 | 75-160 | 18 | 8.0 | 78.6 | - | - | - |
| **5** | 0.016 | 7.0 | 75-140 | 18 | 15.2 | 92.1 | 3.5 | - | - |
| | 2.1 | | | | 13.8 | 90.2 | | | |
| **6** | 2.2 | 7.0 | 75-140 | 18 | 9.7 | 79.4 | - | - | - |
| **7** | 2.2 | 7.0 | 75-160 | **15** | 11.8 | 77.2 | - | - | - |
| **8** | 2.2 | 7.0 | 75-130 | 18 | 11.2 | 68.5 | 4.0 | 68.0 | 2.5 |
| **9** | 2.2 | 7.0 | 75-140 | 18 | 11.6 | 70.3 | 4.0 | 70.5 | 1.8 |
| **10** | 2.2 | 7.0 | 75-160 | 18 | 12.0 | 72.3 | 5.0 | 71.4 | 2.5 |
| **11** | 2.2 | 7.0 | 75-180 | 18 | 12.2 | 65.1 | 5.0 | 67.0 | 2.6 |
| **12** | 0.016 | 7.5 | 75-140 | 18 | 18.6 | 88.3 | 3.9 | - | - |
| | 2.1 | | | | 17.6 | 87.0 | | | |
| **13** | 2.3 | 7.5 | 75-160 | 18 | 16.2 | 72.7 | - | - | - |
| **14** | 0.016 | 8.0 | 75-140 | **15** | 22.5 | 91.7 | 3.2 | - | - |
| | 2.1 | | | | 20.4 | 76.5 | | | |
| **15** | 2.2 | 8.0 | 75-140 | **15** | 19.2 | 81.2 | - | - | - |
| **16** | 2.3 | 8.0 | 75-140 | 18 | 20.0 | 70.4 | - | - | - |

Supplementary Table 1 (Color online) List of characterized devices. From left to right: sequence numbers (No.), nanowire width ($w$) and pitch ($p$); diameter of active area (⌀), switching current ($I_{sw}$), maximal SDE ($SDE_{max}$), polarization extinction rate (PER). $SDE_{ARC}$: maximal SDE after antireflection coating (ARC). $PER_{ARC}$: PER measured after ARC. For the devices (No. 8–11), the influence of the ARC was studied. The ARC layer was deposited by electron-beam evaporation, with a SiO thickness of approximately 408 nm (i.e., $\lambda/2/n_{SiO}$, where $n_{SiO}$ = 1.89 and $\lambda$ is the target wavelength). The value of PER substantially decreased after the ARC, whereas the SDE did not show a notable change.

**Comparison of the designs and performances between NbN-, MoSi-, and WSi-SNSPDs and W-TES**

| SNSPD | | | | | TES |
|---|---|---|---|---|---|
| **Material** | **Nb(Ti)N** | | **MoSi**[1] | **WSi**[2] | **W**[3] |
| **Cavity design** | half-cavity with DBR | double-side cavity | full-cavity with backside Au (Al for TES) mirror | | |
| **SDE (%) approximately 1550 nm** | 92.1@1.8K<br>90.2@2.1K | 80@2.1K[4]<br>76@2.5K[5] | 87.1@0.7K<br>82@2.3K | 93@0.12K<br>90@2K | 95@0.1K |
| **DCR (c/s)** | 10 | 1000 | 100 | 1000 | N/A |
| **Jitter (ps)** | 79@2.1K | 40–80 | 99 | 150@0.12K | 50000 – 100000 |
| **$I_{sw}$ (μA)** | 14.5@1.8K<br>13.8@2.1K | 12@2.1K[4]<br>6.5@2.5K[5] | 9.5@0.7K<br>4.3@2.3K | 4@0.12K<br>1.8@2k | approximately 17@0.1K |
| **Decay time (ns)** | 48.5 | 5[5], approximately 30[4] | 35 | 120 (rest time approximately 40 ns) | 800 |
| **Package** | fiber front-side package | fiber back-side package[4];nano-positioner[5] | front-side fiber self-alignment package | | |

Supplementary Table 2 (Color online) Performance of the Nb(Ti)N, WSi, MoSi-SNSPDs, and W-TES at 1550 nm wavelength.

## Supplementary References